\documentclass[12pt]{article}
\usepackage{a4wide,amsmath,graphicx,psfrag}

\newcommand{\bs}[1]{\boldsymbol{#1}}

\begin{document}
\begin{titlepage}

\begin{flushright}
TTP00-18\\
hep-ph/0008180\\
August 2000\\
\end{flushright}
\vskip 4.0ex
\begin{center}
\boldmath
{\Large\bf Kinematic Effects in \\[0.5ex] Quarkonia Production}
\footnote{Talk given at the conference ``Hyperons, Charm and Beauty Hadrons'',
Valencia, June 2000. To be published in Nucl.~Phys.~Suppl.}
\unboldmath
\vskip 6.0ex
{\sc Stefan Wolf}
\vskip 2.0ex
{\em Institut f\"ur Theoretische Teilchenphysik, Universit\"at Karlsruhe,\\
D-76128 Karlsruhe, Germany}
\end{center}

\begin{abstract}
\noindent
We investigate energy (momentum) distributions in $J/\psi$ photoproduction and
$J/\psi$ production in $B$ meson decay. In particular the upper endpoint region
of the spectrum is examined where the effect of soft gluon emission from the
$c\bar{c}$ pair becomes important. Constructing a model which is consistent
with the so-called shape function formalism we consider these fragmentation
effects and show that the relevance of possible colour octet contributions in
the photoproduction channel is still inconclusive.
\end{abstract}

\vfill

\end{titlepage}

\section{Introduction}
Non-relativistic QCD (NRQCD) \cite{Bodwin:1995jh} provides the framework for a
successful description of many heavy quarkonia production processes. Performing
a systematic expansion in the typical velocity $v$ of a (anti)quark inside the
quarkonium NRQCD includes modes where the quark antiquark pair ($Q\overline{Q}$
pair) is produced in a colour octet state and hadronize to a colourless
quarkonium by soft gluon radiation.

Like other effective theories NRQCD implies a strict factorization between the
short-distance production of the $Q\overline{Q}$ pair and its long-distance
fragmentation into the quarkonium. While one can compute the partonic process
perturbatively to definite order in the strong coupling constant $\alpha_s$ the
non-perturbative physics has to be parameterized into so-called NRQCD matrix
elements $\langle {\cal O}_c^H ({}^{2S+1}\!L_J) \rangle$. They give the
probability for the hadronization of a colour singlet/octet ($c = 1$/$c = 8$)
$Q\overline{Q}$ pair with spin $S$, orbital angular momentum $L$, and total
angular momentum $J$ into the quarkonium $H$ and scale according to power
counting rules of NRQCD, i.e.~for example that the colour octet matrix elements
are suppressed at least by factor of $v^2$ with respect to the leading order
colour singlet contribution.

Although NRQCD matrix elements describe the hadronization adequately in most
quarkonium processes, the theoretical prediction of its energy distributions
needs more effort: Towards the endpoint of the spectrum the quarkonium takes
more and more energy of the incoming particles, i.e.~the phase space for
radiating off soft gluons from the quark antiquark pair becomes less and
less. Therefore one should expect that the colour octet contributions where the
$Q\overline{Q}$ pair must emit a soft gluon to get rid of its colour are
suppressed for large values of the quarkonium energy. However, in the inelastic
$J/\psi$ photoproduction channel naive NRQCD calculations yield a steep raise
of the colour octet contributions in the upper endpoint region
\cite{Cacciari:1996dg} contradicting the observation of a rather flat
spectrum \cite{HERA}. Thus we parameterize the fragmentation process by
functions, the so-called shape functions, rather than by numbers to account for
the phase space effect mentioned above.

Technically speaking the necessity of shape functions instead of matrix
elements is caused by the breakdown of the NRQCD velocity expansion near the
endpoint of quarkonia energy spectra \cite{Beneke:1997qw}. When the quarkonium
carries a fraction ($1 - \epsilon$) of its maximal energy, where $\epsilon$ is
a small number, the kinematic limitations from the hadronic process appear in
the short-distance calculation as expansion in $v^2/\epsilon$. While for small
quarkonia energies ($\epsilon \gg v^2$) the naive use of NRQCD is appropriate
the endpoint region ($\epsilon \sim v^2$) needs a resummation of the leading
twist terms ${\cal O}((v/\epsilon)^k)$. Thus one gets a shape function for each
production channel separately which smears the partonic spectrum. As result the
endpoint moves from the partonic value given by the mass of the $Q\overline{Q}$
pair to the hadronic one set by the quarkonium mass.

The outline of this article is as follows. First we construct a model for the
shape function which is consistent with the shape function formalism of
\cite{Beneke:1997qw} and models the region where $\epsilon \ll v^2$. Thereafter
we apply our model to the decay $B \to J/\psi X$ and fit the model parameter
$\Lambda$ to the CLEO data. Assuming universality we finally transfer the
results from $J/\psi$ production in $B$ decay to the photoproduction
channel.\footnote{For a more detailed presentation of the contents of this
article please refer to \cite{Beneke:2000gq} by Beneke {\it et al.}}

\section{Shape function model}

\begin{figure}[t]
\begin{center}
\includegraphics[width=.6\textwidth]{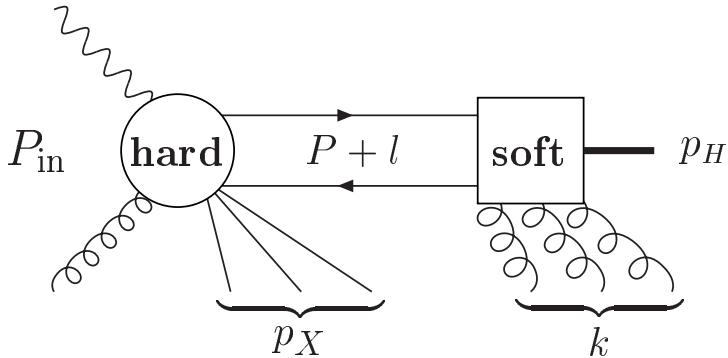}
\end{center}
\caption{Diagrammatic representation of a general quarkonium production process.}
\label{fig1}
\end{figure}

Starting point for our analysis is figure \ref{fig1}. Here quarkonium
production is considered as a two step process: the short-distance production
of the $Q\overline{Q}$ pair described by the partonic cross section
$\hat{\sigma}_{Q\overline{Q}}[n]$ and its subsequent fragmentation into the
quarkonium $H$. Hence we get
\begin{equation}
       (2\pi)^3 \, 2p_H^0 \frac{d\sigma}{d^3p_H}
\equiv \sum_n \int \!\! \frac{d^4l}{(2\pi)^4} \,\,
              \hat{\sigma}_{Q\overline{Q}}[n](l) \cdot F_n(l) 
\end{equation}
where the shape function
\begin{equation}
F_n(l) = \int \!\! \frac{dk^2}{2\pi} \frac{d^3 \bs{k} }{(2\pi)^3 2k_0} \,\,
(2\pi)^4 \delta^4(P + l - p_H - k) \Phi_n(k; p_H, P)
\end{equation}
parameterizes the emission of a soft gluon field with total momentum $\bs{k}$
and invariant mass $k^2$ from a $Q\overline{Q}$ pair of quantum state $n =
{}^{2S+1}\!L_J^{(c)}$ and with off-shellness $l = p_{Q\overline{Q}} - P$ where
$P^2 = 4m_Q^2$. $\Phi_n(k)$ is a radiation function that contains the
non-perturbative physics of the hadronization process. Before defining this
radiation function we elaborate the phase space exactly.

To this end we restrict ourselves to the case of a single massless hard
particle in the final state and refer to the quarkonium rest frame defined by
$\bs{p_H} = \bs{P} = 0$ rather than to the center-of-mass frame ($\bs{P}_{\rm
in} = 0$). After integrating over the hard momentum $p_X$ and the soft momentum
$\bs{k}$ we switch to light cone variables for $l$ to reproduce the shape
function formalism of \cite{Beneke:1997qw}. We integrate out $l_+ = l_0 + l_z$,
$l_\perp^2$ and the azimuthal angular $\phi$ and rewrite the $l_0$ integration
into a integration over the soft gluon energy $k_0$. Then the final result
reads
\begin{equation}
(2\pi)^3 \, 2p_H^0 \frac{d\sigma}{d^3p_H}
= \sum_n \int\limits_0^{\alpha\beta} \frac{dk^2}{2\pi} \!
           \int\limits_{(\alpha^2+k^2)/(2\alpha)}^{(\beta^2+k^2)/(2\beta)}
	                                                     \hspace{-1em} dk_0
	{\cal F} \!\cdot\! \bar{H}_n(P_{\rm in}, P, l, p_X)
                             \cdot \frac{\Phi_n(k; p_H, P)}{4\pi(\beta-\alpha)}
\label{master}
\end{equation}
where the integration bounds are functions of
\begin{equation}
\alpha \equiv P_{\rm in+} - M_H \,,\quad \beta \equiv P_{\rm in-} - M_H
\end{equation}
which again are defined in the quarkonium rest frame.

The hard subprocess $\text{\it in} \to Q\overline{Q}[n] + X$ is given by the
flux factor ${\cal F}$ and the azimuthal average $\bar{H}_n = \int d\phi/(2\pi)
H_n$ of the partonic amplitude squared. While this part can be calculated
perturbatively we have to model the radiation function $\Phi_n$. We make the
ansatz
\begin{equation}
\Phi_n(k; p_H, P) = a_n \cdot |\bs{k}|^{b_n} \exp(-k_0^2/\Lambda_n^2)
                    \cdot k^2\exp(-k^2/\Lambda_n^2)
\end{equation}
where the exponential cut-off reflects our expectation that the typical energy
and invariant mass of the radiated system is of order $\Lambda_n \sim m_Q v^2
\approx$ several hundred MeV for $n={}^1\!S_0^{(8)}$, ${}^3\!P_J^{(8)}$,
${}^3\!S_1^{(8)}$. Note that there is no soft gluon emission necessary in the
colour singlet case, i.e.~$\Phi_n$ collapses to the delta function
$\delta^4(k)$. The normalization $a_n$ is fixed as described in
\cite{Beneke:2000gq}. We choose the other parameters as follows ($c = 1.5$)
\cite{Beneke:2000gq}:
\begin{align}
& b[{}^1\!S_0^{(8)}] = 2 \,, \qquad
        b[{}^3\!P_0^{(8)}] = b[{}^3\!S_1^{(8)}] = 0 \,,
\\
& \Lambda[{}^1\!S_0^{(8)}] = \Lambda[{}^3\!P_0^{(8)}] \equiv \Lambda \,,
\quad \Lambda[{}^3\!S_1^{(8)}] = c \Lambda \,.
\end{align}

\section{Momentum spectrum in $B \to J/\psi X$}

Let us now apply the formalism to the $J/\psi$ momentum spectrum in the
semi-inclusive decay $B \to J/\psi X$. The leading partonic decay process $b
\to c\bar{c}[n] + q$ ($q = \{d, s\}$) results in a $J/\psi$ with fixed momentum
but the hadronic decay spectrum is modified by both initial and final bound
state effects. The Fermi motion of the $b$ quark inside the $B$ meson is
accounted by the ACCMM model in a simple but satisfactory way. The
fragmentation effects are considered by our shape function formalism. Colour
octet contributions are non-negligible because their suppression due to a
scaling factor of $v^4 \approx 1/15$ from the NRQCD matrix elements is
compensated by a factor $C_{[8]}^2/C_{[1]}^2 \approx 15$ from the Wilson
coefficients in the partonic process.

Thus we take the partonic amplitude squared $H_n(m_b, 2m_c)$ for $n =
{}^1\!S_0^{(8)}$, ${}^3\!P_J^{(8)}$, ${}^3\!S_1^{(8)}$, and ${}^3\!S_1^{(1)}$
at tree level \cite{Ko:1996iv} and perform the substitution
\begin{equation}
\label{Mcc}
2m_c \to M_{c\bar{c}}(k) = \sqrt{M_{J/\psi}^2 + 2 M_{J/\psi} k_0 + k^2}
\end{equation}
to allow the $c\bar{c}$ pair being off-shell before it emits soft gluons. Then
the $J/\psi$ energy distribution in $b$ quark decay is obtained by folding the
partonic result with our shape function model. Finally we use this result as
input for the ACCMM model to include initial bound state effects. It assumes an
isotropic Fermi motion of the $b$ quark inside the meson and a Gaussian
momentum distribution with a width $p_F$ of several hundreds MeV. Furthermore
the ACCMM model keeps the kinematics of the $b$ quark decay `in flight' exact
treating the soft degrees of freedom inside the $B$ meson as spectator of mass
$m_{sp}$.

CLEO has measured the $J/\psi$ momentum spectrum on the $\Upsilon(4S)$
resonance \cite{Balest:1995jf}. Hence we boost our result from the $B$ meson
rest frame to their laboratory system. We then assume that the colour singlet
contribution is dual to the exclusive modes $B \to J/\psi K^{(*)}$
\cite{Beneke:2000gq}.

\psfrag{label of x axis}{\hspace{-10.5em}
		\small $J/\psi$ momentum $[\text{GeV}/c]$}
\psfrag{label of y axis}{\hspace{-7.7em}
		\small $d{\rm BR}_{[8]}/dp_{J/\psi}$ $[(\text{GeV}/c)^{-1}]$}
\psfrag{L000}{\small $\Lambda = \;\;\;\,0$ MeV:}
\psfrag{L200}{\small $\Lambda = 200$ MeV:}
\psfrag{L300}{\small $\Lambda = 300$ MeV:}
\psfrag{L500}{\small $\Lambda = 500$ MeV:}
\begin{figure}[t]
\begin{center}
\includegraphics[width=.6\textwidth]{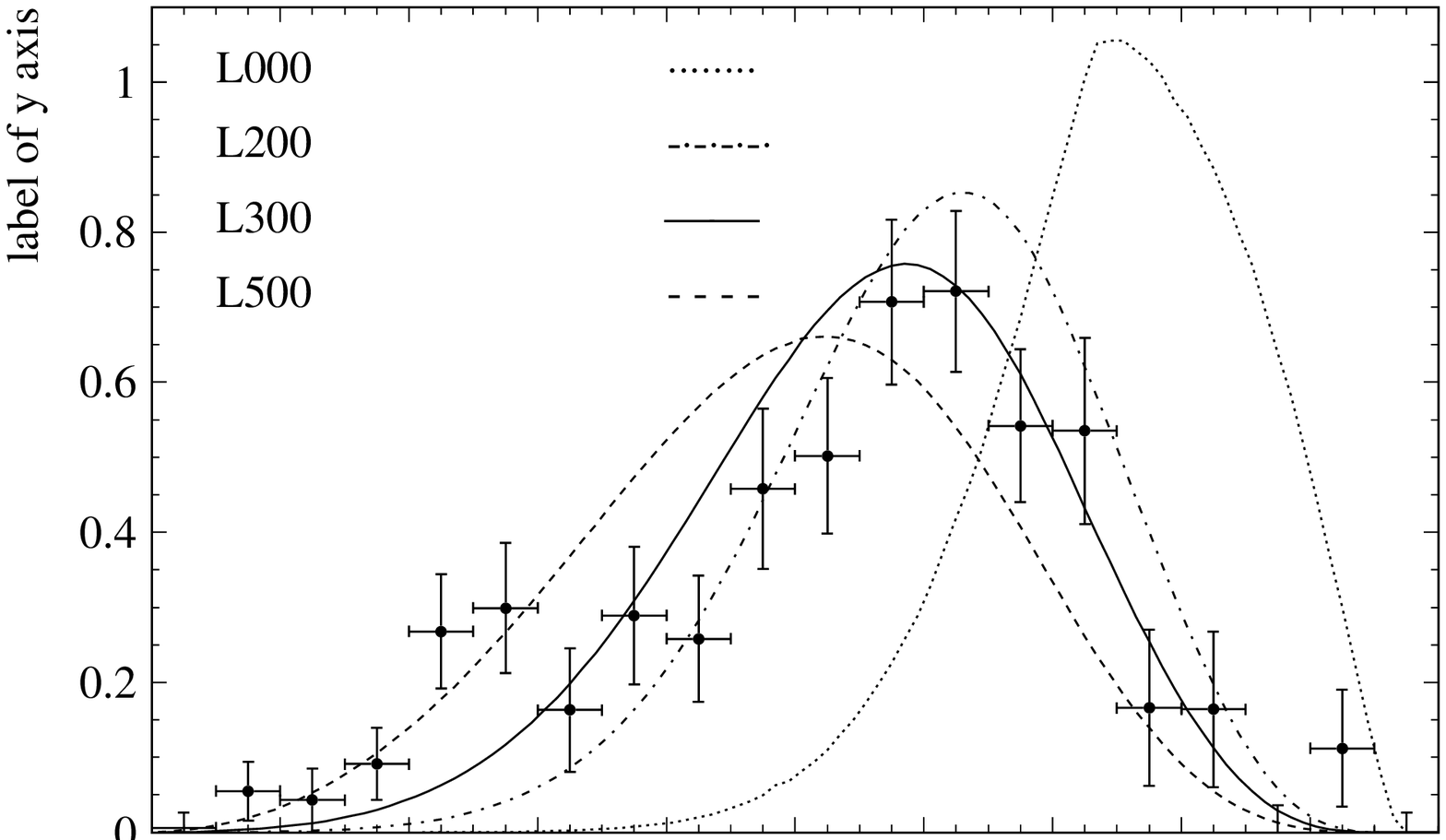}
\end{center}
\vspace{1ex}
\caption{Sum of colour octet modes $d{\rm BR}_{[8]}/dp_R$ to the
         differential branching ratio of the decay $B \to J/\psi X$
         compared to CLEO data \cite{Balest:1995jf}.}
\label{fig2}
\end{figure}

The result for the branching ratio with $J/\psi K^{(*)}$ subtracted is shown
in figure \ref{fig2}. It is clearly seen that the effect of $c\bar{c}$
fragmentation is necessary to reproduce the data for an ordinary choice for the
ACCMM parameters ($p_F = 300$ MeV, $m_{sp} = 150$ MeV). In other words $\Lambda
= 0$ cannot describe the data without pushing $p_F$ up to values far above the
ones successfully employed e.g.~in semi-leptonic $B$ decays. If we fit the
shape function model parameter $\Lambda$ we obtain
\begin{equation}
\label{Lambda}
\Lambda = 300^{+50}_{-100} \, \text{MeV}
\end{equation}
which excellently fits to our expectation $\Lambda \sim m_c v^2 \sim
\Lambda_{\rm QCD}$ from the NRQCD velocity scaling rules. The uncertainties in
(\ref{Lambda}) come from a different weighting between the colour octet
contributions and from varying $p_F$ from 200 MeV to 500 MeV.

\section{Inelastic $J/\psi$ photoproduction} 

In this section we discuss the energy spectrum in inelastic $J/\psi$
photoproduction. Again we take the squared amplitude $H_n$ from the partonic
calculation \cite{Cacciari:1996dg} and perform the substitution (\ref{Mcc})
$2m_c \to M_{c\bar{c}}(k)$.\footnote{There is a misprint in the original paper
\cite{Beneke:2000gq}: In each mode of $H_n$ the factor $g_s^2$ should be
substituted by $g_s^4$. The numerics, however, remain unchanged.} In this case
it is more complicated to incorporate the shape function model (\ref{master})
since the incoming particles distinguish a symmetry axis in the partonic
process $g + \gamma \to c\bar{c}[n] + g$. We will keep exact its perpendicular
$l_\perp$ dependence even though it is of higher order in velocity
scaling. Next we employ the parton density function to include initial state
effect and integrate over the proton momentum fraction of the gluon and the
transverse momentum $p_T$ of the $J/\psi$ to get its energy spectrum
$d\sigma(\gamma p \to J/\psi X)/dz$ where $z = (p_{J/\psi} \!\cdot\!
p_p)/(p_\gamma \!\cdot\! p_p)$ is the photon energy fraction transferred to the
$J/\psi$.

Finally a comment on the normalization of our result. The relative weights of
the different production channels are set by the corresponding values of the
NRQCD matrix elements \cite{Beneke:2000gq}. The overall normalization is fixed
by adjusting the shape function modified curvature to the one from naive NRQCD
in the region $0.1 \le z \le 0.4$. Thus we account for the effect that
partonically the $c$ quark mass (\ref{Mcc}) is higher than the value $m_c =
1.5$ GeV used in the fits for the NRQCD matrix elements.

\psfrag{ii}{$z$}
\psfrag{differential cross section}
	{\hspace{-5em} $d\sigma(\gamma p \to J/\psi X)/dz$ [nb]}
\psfrag{pTcut1}{\small $p_T \ge 1$ GeV}
\psfrag{pTcut2}{\small $p_T \ge 2$ GeV}
\psfrag{pTcut5}{\small $p_T \ge 5$ GeV}
\psfrag{H1}{\small H1:}
\psfrag{94/95}{\small $94/95$}
\psfrag{96/97 prel1.}{\small $96/97$ prel.}
\psfrag{Zeus1}{\small Zeus:}
\psfrag{94}{\small $94$}
\psfrag{97 prel.}{\small $97$ prel.}
\psfrag{Zeus2}{\small Zeus:}
\psfrag{96/97 prel2.}{\small $96/97$ prel.}
\begin{figure}[t]
\begin{center}
\includegraphics[width=.6\textwidth]{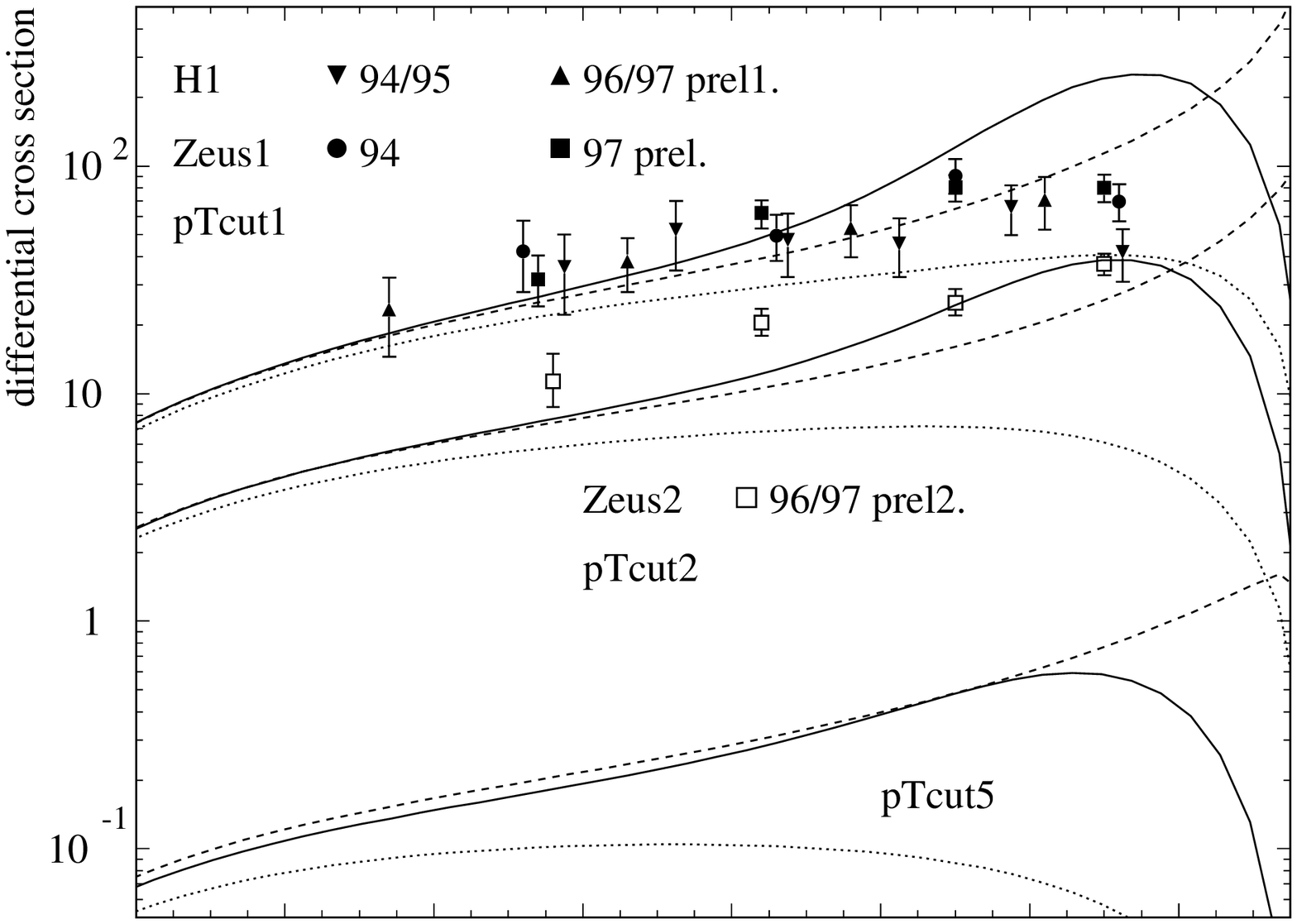}
\end{center}
\caption{$J/\psi$ energy spectrum in photoproduction for different values of
the transverse momentum cut ($p_T^{\rm cut} = 1,2,5$ GeV): colour singlet model
(dotted), naive (dashed) and shape function modified (solid) NRQCD calculation
versus HERA data for $p_T \ge 1$ GeV and $\ge 2$ GeV respectively.}
\label{fig3}
\end{figure}

Figure \ref{fig3} displays the result for $p_T \ge 1$ GeV and $p_T \ge 2$ GeV
plotted against data from HERA \cite{HERA}. Due to our normalization procedure
the shape function improved curves coincide with the naive NRQCD calculation
for low values of $z$. However, they display a strong enhancement in the region
of intermediate $z$ before they fall down to zero for $z \to 1$ as anticipated
from general arguments.

The explanation for this (unphysical) strong enhancement lies in the structure
of partonic colour octet amplitudes. Since $H_n$ is proportional to $(1 -
z_{c\bar{c}})^{-2}$ for $n = {}^1\!S_0^{(8)}$ and $n = {}^3\!P_J^{(8)}$ where
$z_{c\bar{c}}$ is the corresponding quantity to $z$ for the $c\bar{c}$ pair we
get large contributions from $z_{c\bar{c}} \to 1$. Unfortunately soft gluon
radiation in the final state decouples the energy (as well as the transverse
momentum) of the $J/\psi$ and the $c\bar{c}$ pair, i.e.~including our model we
sample the partonic rate for $z_{c\bar{c}} \ge z$ close to one even though the
cut on $p_T$ sets an upper bound on $z$. To fix this problem we repeated the
analysis with a higher $p_T$ cut (cp.~figure \ref{fig3}). Now the hunch
disappears because $z^{\rm max}$ is small enough to keep $z_{c\bar{c}}$ away
from one.

In conclusion one should note that it is impossible to neglect or even rule out
colour octet contributions in the photoproduction channel from the observation
of a flat energy spectrum without increasing $p_T^{\rm cut}$.

\subsection*{Acknowledgements}
I would like to thank M.~Beneke for useful comments and careful reading the
manuscript. The author is supported by the Graduiertenkolleg
``Elementarteilchenphysik an Beschleunigern'' and the DFG-Forschergruppe
``Quantenfeldtheorie, Computeralgebra und Monte-Carlo-Simulation''.

\end{document}